\documentclass[%
 reprint,
superscriptaddress,
 amsmath,amssymb,
 aps,
prb,
]{revtex4-1}

\usepackage{graphicx}
\usepackage{dcolumn}
\usepackage{bm}

\begin{document}


\title{Relaxation dynamics of modulated magnetic phases in skyrmion host GaV$_4$S$_8$: an ac magnetic susceptibility study}

\author{{\'A}d{\'a}m Butykai}
 \email{butykai@dept.phy.bme.hu}
 \affiliation{MTA-BME Lend\"{u}let Magneto-optical Spectroscopy Research Group, 1111 Budapest, Hungary}
\affiliation{Department of Physics, Budapest University of Technology and Economics 1111 Budapest, Hungary}

\author{S{\'a}ndor Bord{\'a}cs}%
\affiliation{MTA-BME Lend\"{u}let Magneto-optical Spectroscopy Research Group, 1111 Budapest, Hungary}
\affiliation{Department of Physics, Budapest University of Technology and Economics 1111 Budapest, Hungary}

\author{L{\'a}szl{\'o} F. Kiss}
\affiliation{Department of Experimental Solid State Physics, Institute for Solid State Physics and Optics, Wigner-MTA Research Centre for Physics, 1121 Budapest, Hungary}

\author{Bertalan Gy\"{o}rgy Szigeti}%
\affiliation{Department of Physics, Budapest University of Technology and Economics 1111 Budapest, Hungary}

\author{Vladimir Tsurkan}
\affiliation{Experimental Physics V, Center for Electronic Correlations and Magnetism, University of Augsburg, 86135 Augsburg, Germany}
\affiliation{Institute of Applied Physics, Academy of Sciences of Moldova, MD 2028, Chisinau, Republica Moldova}

\author{Alois Loidl}
\affiliation{Experimental Physics V, Center for Electronic Correlations and Magnetism, University of Augsburg, 86135 Augsburg, Germany}

\author{Istv{\'a}n K{\'e}zsm{\'a}rki}
\affiliation{MTA-BME Lend\"{u}let Magneto-optical Spectroscopy Research Group, 1111 Budapest, Hungary}
\affiliation{Department of Physics, Budapest University of Technology and Economics 1111 Budapest, Hungary}
\affiliation{Experimental Physics V, Center for Electronic Correlations and Magnetism, University of Augsburg, 86135 Augsburg, Germany}

\date{\today}

\begin{abstract}
We report on the slow magnetization dynamics observed upon the magnetic phase transitions of GaV$_4$S$_8$, a multiferroic compound featuring a long-ranged cycloidal magnetic order and a N{\'e}el-type skyrmion lattice in a relatively broad temperature range below its Curie temperature. The fundamental difference between GaV$_4$S$_8$ and the chiral helimagnets, wherein the skyrmion phase was first discovered, lies within the polar symmetry of GaV$_4$S$_8$, promoting a cycloidal instead of a helical magnetic order and rendering the magnetic phase diagram essentially different from that in the cubic helimagnets. Our ac magnetic susceptibility study reveals slow relaxation dynamics at the field-driven phase transitions between the cycloidal, skyrmion lattice and field-polarized states. At each phase boundary, the characteristic relaxation times were found to exhibit a strong temperature dependence, starting from the minute range at low temperatures, decreasing to the micro- to millisecond range at higher temperatures.

\end{abstract}

\pacs{Valid PACS appear here}
\maketitle


\section{\label{sec:Intro}Introduction}

Magnetic skyrmions are topologically non-trivial, whirling spin structures, which can form 2-dimensional crystals, so-called skyrmion lattices \cite{bogdanov1989thermodynamically,nagaosa2013topological}. The emergence of the skyrmion lattice phase was first identified in the close vicinity of the paramagnetic-helical phase boundary of cubic chiral helimagnets, known as B20 compounds with a P2$_1$3 space group \cite{muhlbauer2009skyrmion,yu2010real,uchida2008topological,yu2011near,shibata2013towards}. Cu$_2$OSeO$_3$, belonging to the same space group with a different crystal structure \cite{meunier1976constantes} was the first insulating material demonstrated to host skyrmions \cite{adams2012long,seki2012observation}, with a magnetoelectric character \cite{white2012electric,seki2012magnetoelectric,white2014electric,liu2013skyrmion}. Since their experimental discovery, skyrmions have attracted much attention owing to their potential application as magnetic bits in high-capacity and low-consumption memory devices \cite{jonietz2010spin,fert2013skyrmions,koshibae2015memory,zhang2015magnetic,zhang2015skyrmion}.

GaV$_4$S$_8$, a member of the lacunar spinel family, characterized by the space group F$\bar{4}$3m, is the first known example of skyrmion-host materials with non-chiral but polar crystal structure \cite{kezsmarki2015neel}. GaV$_4$S$_8$ is a semiconductor with a non-centrosymmetric cubic crystal structure (T$_d$) at room temperature. The compound undergoes a cooperative Jahn-Teller distortion at T$_S$=42\,K through the stretching of the lattice along any of the four cubic body diagonals reducing the crystal symmetry to polar rhombohedral \cite{pocha2000electronic, muller2006magnetic,wang2015polar,hlinka2016lattice}. As a result a sizable ferroelectric polarization develops along the C$_{3v}$ rhombohedral axis  \cite{ruff2015multiferroicity,xu2015unusual}. Depolarization field is reduced by the formation of submicron-sized structural domains of the four possible rhombohedrally distorted variants with the different $\left< 111 \right>$-type rhombohedral axes \cite{pocha2000electronic}, assembling into an alternating lamellar domain structure \cite{butykai2017characteristics}.

Long-range magnetic ordering arises and thereby the compound becomes a type-I multiferroic at T$_C$=13\,K \cite{yadav2008thermodynamic,ruff2015multiferroicity,xu2015unusual,widmann2016multiferroic}. The interplay of the symmetric exchange interaction and the antisymmetric Dzyaloshinskii-Moriya (DM) exchange interaction gives rise to a long-wavelength spin ordering. However, in contrast to the B20 compounds featuring a helical spin order in zero field and longitudinal conical spin structure in finite magnetic fields, in GaV$_4$S$_8$ the different pattern of DM vectors, dictated by the polar C$_{3v}$ symmetry, leads to a cycloidal spin order (Cyc) and precludes the emergence of the longitudinal conical structure in finite fields \cite{kezsmarki2015neel}. The cycloidal nature of the magnetic modulations has been confirmed experimentally by polarized SANS measurements \cite{bordacs2017sans}.

In low magnetic fields a N{\'e}el-type skyrmion lattice (SkL) develops. Due to the lack of the longitudinal conical state, being the main competitor of the SkL phase in the B20 compounds, the N{\'e}el-type SkL is stable over a broad temperature region \cite{kezsmarki2015neel}. On the other hand, the easy-axis anisotropy in the rhombohedral domains \cite{ehlers2016exchange} promotes the ferromagnetic ordering (FM) of the spins in the ground state, suppressing the modulated phases below T=5\,K \cite{kezsmarki2015neel}. 

The modulation vectors of the Cyc and SkL states are confined to the rhombohedral plane, irrespective of the direction of the applied magnetic field \cite{kezsmarki2015neel}. Furthermore, the skyrmions were demonstrated to carry magnetoelectric polarization \cite{ruff2015multiferroicity,widmann2016multiferroic}, offering new possibilities of their manipulation with external electric fields. 


Measurement of static magnetization or magnetic susceptibility is a primary methodology to reveal the phase transitions between the modulated magnetic states in non-centrosymmetric magnets \cite{bauer2016generic, bauer2010quantum, adams2012long, nakamura2009low, kezsmarki2015neel}. Beyond the determination of the phase diagram via dc susceptibility measurements, recently, several works have been devoted to analyzing the dynamic response of the modulated spin structures via ac susceptibility measurements. Similar features have been identified in the ac susceptibility in many compounds of the B20 family as well as in CuO$_2$SeO$_3$ \cite{bauer2016generic,bauer2012magnetic,levatic2014dissipation,ou2015dynamic, bannenberg2016magnetic,qian2016phase}. Generally, the real component of the ac susceptibility follows well the static susceptibility, however in the vicinity of the magnetic phase boundaries, it deviates from the static value and becomes strongly frequency dependent, accompanied by a finite imaginary component of the susceptibility. In chiral helimagnets, this feature is generally interpreted as the signature of first-order transitions between the different magnetic phases, involving slow relaxation processes of large correlated magnetic volumes, such as the reorientation of the long-wavelength spin-spirals or the nucleation of skyrmionic cores \cite{bauer2012magnetic,levatic2014dissipation,bannenberg2016magnetic,qian2016phase}. The analysis of the frequency-dependence of the ac susceptibility in all cases revealed a broad distribution of relaxation times with macroscopic values at the phase boundaries \cite{levatic2014dissipation,bannenberg2016magnetic,qian2016phase}.  

Here, we report a systematic study of the ac susceptibility in GaV$_4$S$_8$, a compound essentially different from the cubic helimagnets in terms of crystal symmetries and magnetic phase diagram \cite{kezsmarki2015neel}. The analysis of the susceptibility measurements reveals a dramatic increase of the relaxation times in the vicinity of the magnetic phase boundaries, with a strong temperature dependence, which is similar to findings in other systems \cite{levatic2014dissipation,bannenberg2016magnetic,qian2016phase}. In magnetic fields close to the critical values, the average relaxation times are much shorter than 1\,ms at the high-temperature end of each phase boundary and rise well above the minute scale at the low-temperature limits. This strong temperature dependence invokes a thermally activated behavior of the relaxation processes characterized by energy barriers in the order of 1000\,K. 



\section{Relaxation model}

The modulated magnetic structures in skyrmion host compounds are generally characterized by a long correlation length, forming coherent magnetic regions with dimensions of hundreds of nanometers \cite{muhlbauer2009skyrmion,kezsmarki2015neel}. The ac susceptibility of correlated spin-structures consisting of clusters and/or domains of various volumes is generally described by the Cole-Cole relaxation model. This phenomenological model has been effectively applied to various systems \cite{balanda2013ac} comprising large magnetic volumes, such as spin glasses \cite{mulder1981susceptibility}, superparamagnetic nanoparticles \cite{luis1999resonant}, and more recently it has been proposed for the description of the phase transitions between modulated magnetic states in chiral helimagnets, Fe$_{1-x}$Co$_x$Si \cite{bannenberg2016magnetic} and Cu$_2$OSeO$_3$ \cite{levatic2014dissipation,qian2016phase}. 

The dynamic response of the magnetic system in the Cole-Cole model is formulated as an extension of the Debye-relaxation by introducing a distribution of relaxation times, while keeping the exponential time dependence of the relaxation \cite{huser1986phenomenological, balanda2013ac}:

\begin{equation}
\chi^{\omega}=\chi_{\infty}+(\chi_{0}-\chi_{\infty})\frac{1}{1+(i\omega\tau_{c})^{1-\alpha}},
\label{Eq:complex}
\end{equation}

where $\tau_c$ represents the central value of the relaxation times and the $\alpha$ parameter is connected to the width of their distribution. The adiabatic susceptibility, $\chi_{\infty}$ originates from the almost immediate response of the spins, as compared to the time scale of the studied relaxation processes, hence, it is a purely real quantity. The static limit of the ac susceptibility is denoted as $\chi_{0}$. The distribution of the relaxation times, g$(\ln(\tau))$, is expressed by the $\tau_c$ and $\alpha$ parameters as follows \cite{huser1986phenomenological}:

\begin{equation}
g(\ln{\tau})=\frac{1}{2\pi}\left(\frac{\sin{\alpha\pi}}{\cosh{[(1-\alpha)\ln{(\tau/\tau_c)}-\cos{\alpha\pi}]}} \right).
\label{Eq:tau}
\end{equation}

The distribution is symmetric on the logarithmic scale with the central value of $\tau_c$. Zero value of $\alpha$ represents a single Debye-relaxation process, while values close to unity lead to an infinitely broad distribution of relaxation times. 

Owing to the phase sensitivity of the ac-susceptibility measurements, both the real and imaginary components of the susceptibility, $\chi'$ and $\chi''$, can be recovered. The frequency dependence of the two components, expressed from Eq.\ref{Eq:complex}, reads as:

\begin{equation}
\chi'=\chi_{\infty}+(\chi_{0}-\chi_{\infty})\frac{\omega\tau_{c}^{\alpha-1}+\sin\left(\frac{\alpha\pi}{2}\right)}{\omega\tau_{c}^{\alpha-1}+\omega\tau_{c}^{1-\alpha}+2\sin\left(\frac{\alpha\pi}{2}\right)}
\label{Eq:real}
\end{equation}

\begin{equation}
\chi''=(\chi_{0}-\chi_{\infty})\frac{\cos\left(\frac{\alpha\pi}{2}\right)}{\omega\tau_{c}^{\alpha-1}+\omega\tau_{c}^{1-\alpha}+2\sin\left(\frac{\alpha\pi}{2}\right)}.
\label{Eq:imag}
\end{equation}

\section{\label{sec:Methods}Methods}
\subsection{Sample synthesis and characterization}
Single crystalline GaV$_4$S$_8$ samples were grown by chemical vapour transport method using iodine as transport agent. The high crystalline quality of the samples has been confirmed by X-ray diffraction. A cuboid-shaped sample with a mass of 23.4\,mg was selected for the ac susceptibility measurements. 

\subsection{Static and ac susceptibility measurements}
A 5T Quantum Design MPMS SQUID magnetometer was used for the static and ac susceptibility measurements. Both the static magnetic field and the ac modulation field were normal to the (111) plane of the GaV$_4$S$_8$ crystal and the longitudinal magnetic moment was measured in a phase sensitive manner. The field dependence of the ac susceptibility was measured in the 0-80\,mT range with a modulation amplitude of $\mu_0H^{\omega}$=0.3\,mT. In order to probe the dynamics of the magnetically ordered spin system in the low-frequency regime, the drive frequency of the modulating coil was varied between f=0.1\,Hz and 1\,kHz. The in-phase and out-of-phase components of the oscillating magnetization were measured and normalized by the drive amplitude, $\mu_0$H$^{\omega}$ to obtain the real and imaginary parts of the ac-susceptibility, respectively. The dc magnetization was measured in a subsequent run after the ac susceptibility measurements. The static susceptibility, $\partial m/\partial H$, was obtained from the measured dc magnetization curves by numeric derivation using the central difference method. The typical duration of the static measurements was $\approx$100\,s per data point, which involved the ramping of the magnetic field and performing the measurement by the reciprocating sample option (RSO). 

\section{\label{sec:Results}Results and discussion}

\begin{figure*}
    \centering
    \includegraphics[width=\textwidth]{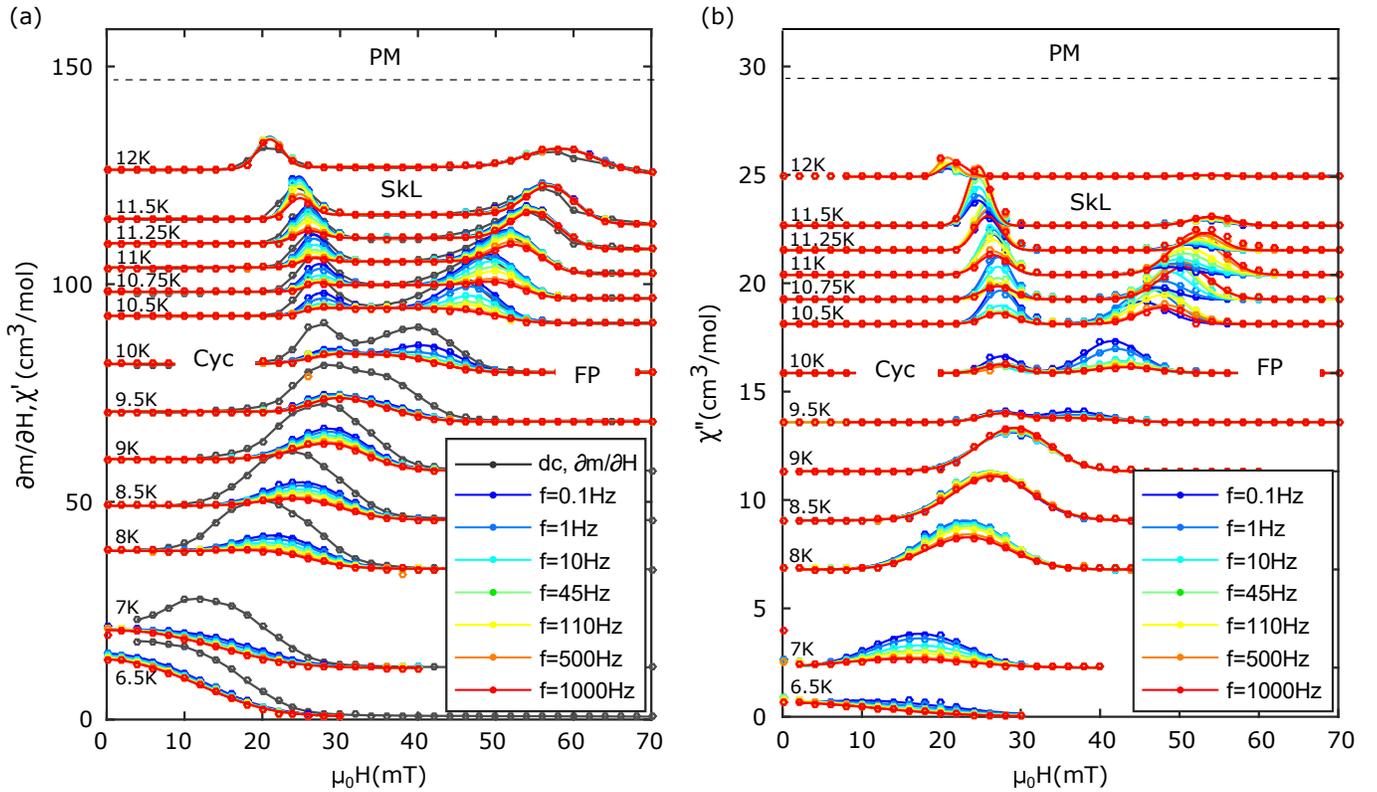}
    \caption{(Color online) Static and ac susceptibility measured in GaV$_4$S$_8$, plotted against the external magnetic field. Panel (a) presents the static susceptibility, $\partial m/\partial H$, plotted in gray, along with the real component of the ac susceptibility, $\chi'(H)$. Panel (b) displays the imaginary component of the ac susceptibility, $\chi''(H)$. Different colors represent data measured at various ac frequencies in the f=0.1\,Hz-1\,kHz range. Measured data are shifted proportionally with the sample temperature, which is indicated on the left side of each curve. The continuous lines connecting the dots are guides to the eye. The magnetic phases separated by the susceptibility peaks in the B-T plane are indicated in the graphs. The paramagnetic state above T$_C$=13\,K is denoted as PM.}
    \label{fig:chi_abs}
\end{figure*}

The magnetic-field dependence of the static and ac susceptibility was recorded at various temperatures within the magnetically ordered phases of GaV$_4$S$_8$, i.e.~between T=6.5\,K and T=12\,K with magnetic fields applied parallel to the crystallographic $\left< 111 \right>$-axis. The static susceptibility and the real part of the ac susceptibility are shown in Fig. \ref{fig:chi_abs} (a). The imaginary component of the ac susceptibility is presented in Panel (b) in the same figure. 

\subsection{Magnetic phase diagram established by the static susceptibility}
The field-driven phase transitions are associated with peaks in the field dependence of the static susceptibility \cite{bauer2016generic,kezsmarki2015neel}. The susceptibility measurements in Fig.\ref{fig:chi_abs} reveal the same magnetic phase diagram as established by K{\'e}zsm{\'a}rki \textit{et al.} in earlier static magnetization measurements \cite{kezsmarki2015neel}. The peaks in the dc susceptibility separate three magnetic phases in the temperature -- magnetic field plane identified as the Cyc, SkL and FP states. The three states adjoin at the triple point located at T=9.5\,K and $\mu_0H$=32\,mT. Despite the coexistence of structural domains in the compound, within the studied field range, phase transitions are only observed in the structural variant wherein the rhombohedral axis is parallel to the field \cite{kezsmarki2015neel}. Remarkably, the susceptibility is enhanced in the SkL phase by almost a factor of 2 as compared to the Cyc phase, implying a larger susceptibility of the former phase along the rhombohedral axis. In the FP state the susceptibility is nearly zero. 

\subsection{Slow relaxation phenomena at the magnetic phase boundaries}
In the following, we discuss the ac susceptibility measurements, indicating slow dynamic processes occurring in the modulated magnetic phases of GaV$_4$S$_8$.

It is clearly seen in Figs. \ref{fig:chi_abs} (a) and (b) that away from the phase transitions, the ac susceptibility is frequency-independent and purely real, with values identical to the static susceptibility, i.e.~they correspond to $\chi_{\infty}$. In these regions, the characteristic relaxation times are much shorter than 1\,ms, the time period of the highest-frequency modulation in our experiments. 
 
On the other hand, as approaching the magnetic phase boundaries, the real component of the susceptibility falls behind the static values, exhibiting a strong frequency dependence. This effect becomes the most prominent below T=9.5\,K, i.e.~for the Cyc-FP phase transition. Moreover, near each phase transition, a peak is seen the imaginary component of the ac susceptibility, as a signature of dissipative processes occurring at extremely low frequencies, which also accounts for the frequency dependence of both components of the ac susceptibility. Such behavior was also reported in cubic helimagnets of the B20 family \cite{bauer2010quantum,wilhelm2012confinement,bauer2012magnetic,bannenberg2016magnetic} and in Cu$_2$OSeO$_3$ \cite{levatic2014dissipation,qian2016phase} near the magnetic phase transitions. 

A significant difference is that in the cubic helimagnets, the spiral magnetic order undergoes multiple phase transitions. In the absence of a magnetic field, a multidomain helical state is realized with propagation vectors selected by the cubic anisotropy in these materials \cite{ishikawa1976helical,bak1980theory,lebech1989magnetic,adams2012long,seki2012observation}. As the magnetic field is increased and the weak anisotropy is overcome by the Zeeman energy, the domains redistribute upon a first-order phase transition \cite{bauer2017symmetry}. Finally, the wave vectors of the modulations flip towards the magnetic field, establishing a mono-domain longitudinal conical structure. Ac susceptibility measurements in FeGe \cite{wilhelm2012confinement}, MnSi \cite{bauer2016generic,bauer2017symmetry}, Cu$_2$OSeO$_3$ \cite{levatic2014dissipation,qian2016phase} and Fe$_{1-x}$Co$_{x}$Si \cite{bannenberg2016magnetic} indicate that these transitions occur on macroscopic time scales, attributed to the rearrangement of large magnetic spirals.


In GaV$_4$S$_8$, on the other hand, no trace of slow dynamics is seen in magnetic fields away from the Cyc-FP, SkL-FP and Cyc-SkL phase boundaries, even though multiple dynamic processes are expected within the Cyc as well as the SkL phase. According to theoretical considerations generic to spin-spirals in external magnetic field \cite{izyumov1984modulated} and SANS measurements in particular in GaV$_4$S$_8$ \cite{kezsmarki2015neel}, besides the anharmonic deformation of the cycloids and skyrmions induced by the field, the cycloidal wavelength and the skyrmion lattice constant increase substantially with increasing magnetic fields. The absence of dissipation within the Cyc and SkL phases suggests that this expansion takes place on much faster time-scales than 1\,ms in both phases. Furthermore, as opposed to the cubic helimagnets, there is no sign of an abrupt redistribution of the cycloidal wave vectors induced by the field.

Our susceptibility data indicate slow dynamics only near the phase boundaries between the Cyc-FP, SkL-FP and Cyc-SkL phases. At the two boundaries of the SkL phase, the first-order nucleation processes of the skyrmions may account for the long time scales. As opposed to the transition from the longitudinal conical phase to the field polarized state in the B20s and Cu$_2$OSeO$_3$, dissipative processes occur at low frequencies at the Cyc-FP transition in GaV$_4$S$_8$. This suggests the first-order character of the transition, further supported by recent neutron scattering experiments \cite{bordacs2017sans}. The underlying dynamics may be related to the unwinding of the 2$\pi$ domain walls of the anharmonic spin-cycloids polarized by the external field.


\subsection{Temperature and magnetic-field dependence of the relaxation processes}

The frequency dependence of the peaks in both susceptibility components [Fig. \ref{fig:chi_abs} (a) and (b)] shows a strong variation with the temperature. Concerning the Cyc-FP phase boundary, at T=7\,K the dissipation is the largest for the smallest modulation frequency, whereas at T=9\,K it becomes almost frequency independent. Above the triple point between T=10\,K and T=11\,K, a reversal can be seen in the hierarchy of the frequencies in the dissipation peaks at both the Cyc-SkL and SkL-FP phase boundaries, indicating that the characteristic frequencies of the relaxation processes pass through the measurement window. Additionally, the peaks in the imaginary part of the susceptibility are shifted in magnetic field with the change of the drive frequency, which is most prominent at T=10.5\,K. In order to systematically investigate the behavior of the relaxation as the function of the temperature and magnetic field, the frequency dependence of the real and imaginary components of the susceptibility was analyzed at all measured (H,T) points in the vicinity of the magnetic phase boundaries.

\begin{figure}
    \centering
    \includegraphics[width=\columnwidth]{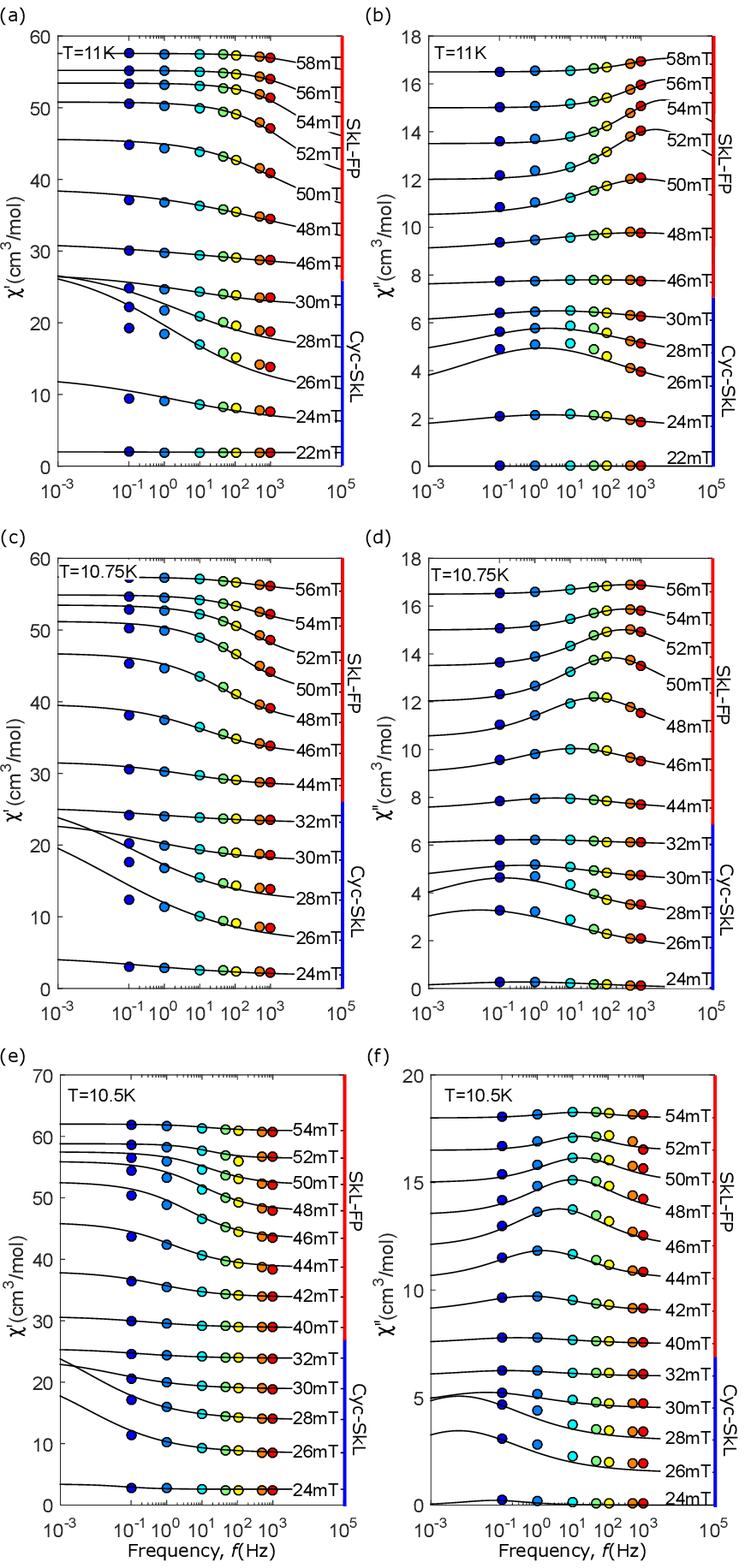}
    \caption{(Color online) Frequency dependence of the real (left column) and imaginary components (right column) of the susceptibility in various magnetic fields above the temperature of the triple point, at T=11K (top), T=10.75K (middle) and T=10.5K (bottom row). Measured values are shifted with respect to the magnetic field. The color coding of the measured values represents the different ac frequencies in accordance with Fig. \ref{fig:chi_abs}. Solid lines are fitted curves according to Eqs. \ref{Eq:real} and \ref{Eq:imag} with the same set of parameters for the real and imaginary parts of the susceptibility. The blue and red bars next to the right axes represent the range of magnetic fields corresponding to the Cyc-SkL and SkL-FP phase transitions, respectively.}
    \label{fig:fits_highT}
\end{figure}

Figure \ref{fig:fits_highT} presents the frequency dependence of $\chi'$ and $\chi''$ in various magnetic fields. Three representative temperatures are selected above the triple point, where the peak in the imaginary part of the susceptibility passes over the experimental window, indicating that the inverse relaxation times go through the range of the measurement frequencies. Figure \ref{fig:fits_highT} only shows data measured in representative magnetic field regions near the Cyc-SkL and SkL-FP phase transitions. The frequency dependence of the complex susceptibility can be fitted well by the Cole-Cole relaxation model (Eqs. \ref{Eq:real} and \ref{Eq:imag}) using the same set of parameters for the real and imaginary components. The shifting of the peak in $\chi''$ towards lower frequencies with decreasing temperature is well traced by the fitted curves for both the Cyc-SkL and the SkL-FP transitions implying an overall slowing down of the relaxation.

\begin{figure*}
    \centering
    \includegraphics[width=\textwidth]{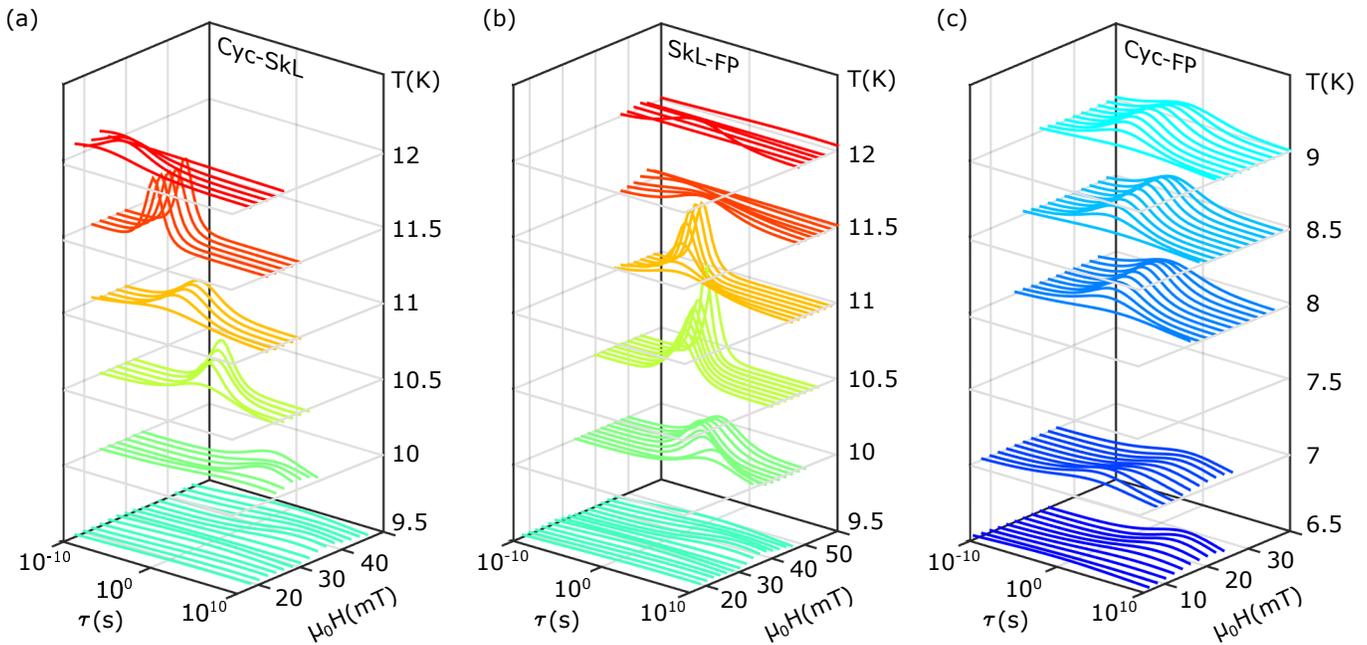}
    \caption{(Color online) Distribution of relaxation times, $g(\ln{(\tau)})$, plotted as a function of the temperature and magnetic field. The distributions were calculated employing Eq. \ref{Eq:tau} with the $\tau_c$ and $\alpha$ parameters obtained from the fits to the frequency dependence of the complex susceptibility. Panels (a),(b) and (c) present the relaxation times in the ranges of magnetic fields corresponding to the Cyc-SkL, SkL-FP and Cyc-FP transitions, respectively. The distribution curves are shifted proportionally with the temperature along the z-axis, which is also indicated in the right side of the graphs. The curves are colored according to a color map representing decreasing temperatures ranging from T=12\,K (red) to T=6.5\,K (blue). The average relaxation time shows a dramatic decrease with decreasing temperatures along all the three phase boundaries, with values well above 10\,s at their low-temperature limits.} 
    \label{fig:tau}
\end{figure*}

Using the $\tau_c$ and $\alpha$ parameters retrieved from the Cole-Cole fits, the distribution of the relaxation times, $g(\ln{\tau})$ was calculated for each (H,T) point, according to Eq. \ref{Eq:tau}. Figures \ref{fig:tau} (a), (b) and (c) display the calculated distributions of the relaxation times for the Cyc-SkL, SkL-FP and Cyc-FP phase transitions, respectively. For each transition, the characteristic time scales fall below $\tau<<1$\,ms at the high-temperature end of the phase boundary, exhibiting a further dramatic increase towards lower temperatures, reaching values $>>$10\,s at the low-temperature part of the phase boundaries. Similar tendencies have been identified in Cu$_2$OSeO$_3$ \cite{levatic2014dissipation,qian2016phase}.


\begin{figure}
    \centering
    \includegraphics[width=\columnwidth]{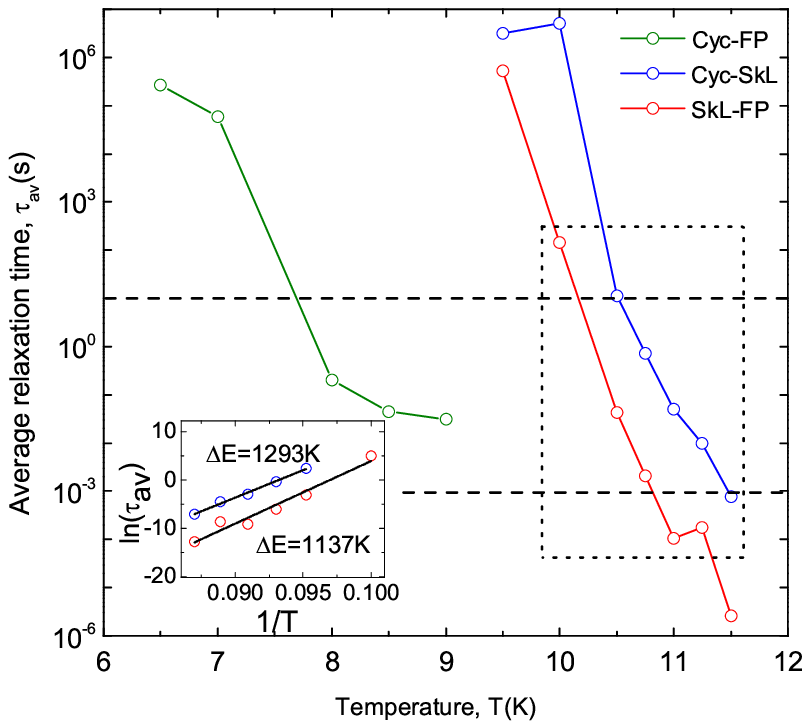}
    \caption{(Color online) Temperature dependence of the logarithmic average of the relaxation times obtained from the Cole-Cole fits, where the averaging was performed over the fitted values in the magnetic field region close to the phase transitions. The green, red and blue circles correspond to the average relaxation times at the Cyc-FP, Cyc-SkL and SkL-FP transitions. The lines connecting the data points are guides to the eye. The dashed horizontal lines represent the measurement window defined by the inverse of the highest (1\,kHz) and lowest (0.1\,Hz) ac frequencies. The inset presents ~$\ln{\tau_{av}}$ as the function of $1/T$ along the Cyc-SkL and SkL-FP phase boundaries. Relaxation time values close to the experimental window are plotted, as indicated by the black dotted frame. Linear fits to the data (solid black lines) yield the average activation energies of 1293\,K and 1137\,K for the Cyc-SkL and SkL-FP transitions, respectively.} 
    \label{fig:tau_vs_T}
\end{figure}

Figure \ref{fig:tau_vs_T} presents the temperature dependence of the fitted relaxation times averaged over the range of magnetic fields near the phase transitions as $\log{(\tau_{av})}=\sum_i\log{(\tau(H_i))}$. The sum runs over the values of relaxation times, $\tau(H_i)$, which are determined by fitting at each field, $H_i$, where the susceptibility shows observable frequency dependence in the vicinity of the phase boundaries. The rapid drop in the relaxation times with increasing temperatures is clearly seen for each phase boundary. The discontinuous jump in the relaxation time at the triple point marks an abrupt change in the relaxation processes between the Cyc-FP and the Cyc-SkL phase. 

The exponential character of the temperature dependence of the relaxation times suggests a thermally activated behavior related to an energy barrier, $\Delta E$, separating the two thermodynamically stable magnetic phases. The energy barriers for the Cyc-SkL and SkL-FP phase transitions were determined by linear fits to the Arrhenius-plots, i.e.~$\ln{(\tau_{av})}$ against $1/T$, as presented in the inset of Fig. \ref{fig:tau_vs_T}. The fitted values yield average activation energies of 1293\,K and 1137\,K at the Cyc-SkL and the SkL-FP boundaries, respectively. These large values underline the stability of the modulated phases extending over large volumes, i.e.~fluctuations of sizable regions instead of individual spins. Since the susceptibility at the Cyc-FP boundary could not be accurately fitted (as discussed later), the relaxation times for this transition have not been analyzed quantitatively. 

Apparently, the characteristic relaxation times are strongly affected also by the magnetic field in case of the SkL-FP transition [Fig. \ref{fig:tau} (b)], whereas such prominent field dependence is not found at the other two phase transitions [Figs. \ref{fig:tau} (a) and (c)]. In the former case, the relaxation substantially accelerates with increasing magnetic field, as approaching the FP state. This may be explained by a phase coexistence of the SkL and FP states with a decreasing typical size of persisting SkL islands towards larger magnetic fields, exhibiting a faster response to the ac modulation. 

\begin{figure}
    \centering
    \includegraphics[width=\columnwidth]{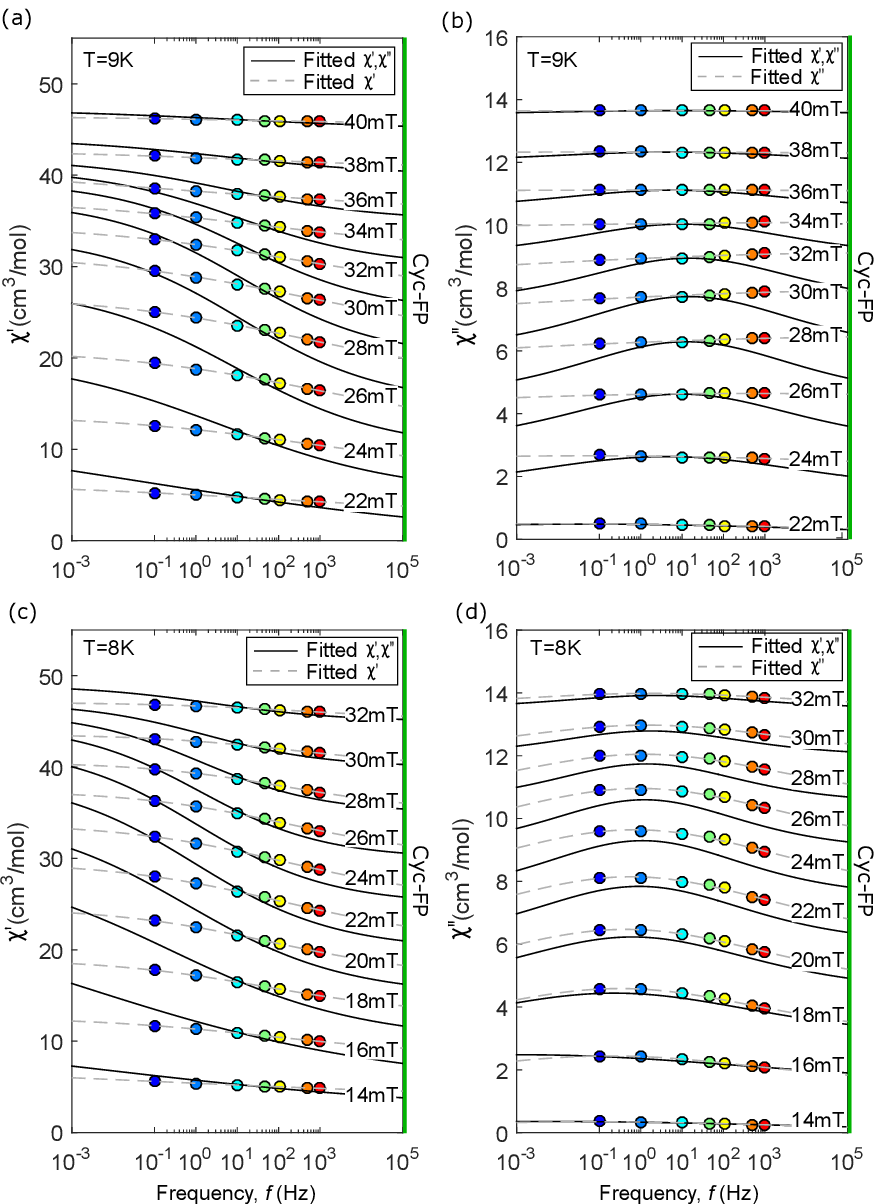}
    \caption{(Color online) Frequency dependence of the real (left column) and imaginary components (right column) of the susceptibility in various magnetic fields below the temperature of the triple point at T=9K (top), T=8K (bottom row). Measured values are shifted with respect to the magnetic field. The color coding of the measured values represents the different ac frequencies in accordance with Fig. \ref{fig:chi_abs}. Solid lines are fitted curves according to Eqs. \ref{Eq:real} and \ref{Eq:imag} with the same set of parameters for the real and imaginary parts of the susceptibility. Gray dashed lines are separate fits to the real [(a) and (c)] and the imaginary components [(b) and (d)] of the susceptibility. The green bars next to the right axes emphasize that the range of magnetic fields at the given temperatures correspond to the Cyc-FP phase transition.}
    \label{fig:fits_lowT}
\end{figure}

In contrast to the other two phase boundaries, the Cole-Cole model fails to fit the frequency dependence of the complex susceptibility for the Cyc-FP transition, as demonstrated in Fig. \ref{fig:fits_lowT} for two selected temperatures below the triple point. Even though the real and the imaginary components can be fitted separately with two different sets of parameters (see dashed gray curves in Fig. \ref{fig:fits_lowT}), the resulting parameters convey no physical meaning, as the Kramers-Kronig relation does not hold between the two components of the response function. The large difference between the static susceptibility values and the real part of the ac susceptibility measured even at the lowest frequency of f=0.1\,Hz, as seen in \ref{fig:chi_abs} (a), suggests that dynamic processes exist with characteristic relaxation times far beyond 10\,s. 

The Cole-Cole model assumes a symmetric distribution of relaxation times on the logarithmic scale \cite{huser1986phenomenological}, which may not apply for more complex processes involved in the magnetic phase transitions in GaV$_4$S$_8$. A generalization of the Cole-Cole function was provided by Havriliak and Negami \cite{havriliak1966complex} allowing for an asymmetric distribution of relaxation times \cite{zorn1999applicability}. Applying the Havriliak-Negami model to our data, however, yielded the same parameters as the Cole-Cole fits returning the same symmetric distribution of relaxation times, hence did not improve the fit. 


Only a few recent studies made an attempt to quantitatively describe the relaxation processes at the magnetic phase boundaries in cubic skyrmion host compounds, each within the framework of the Cole-Cole model \cite{levatic2014dissipation,bannenberg2016magnetic,qian2016phase}. However, in most of these studies the real and imaginary components of the ac susceptibility were handled separately, which may lead to unphysical parameters, as seen for the Cyc-FP transition in GaV$_4$S$_8$ (Fig. \ref{fig:fits_lowT}). Qian \textit{et al.} correlated the Cole-Cole fits to the real and the imaginary parts of the susceptibility in Cu$_2$OSeO$_3$, finding good agreement in case of the conical-to-skyrmion and skyrmion-to-conical transitions, whereas a discrepancy was reported at the helical-to-conical transition. The authors attributed this difference to additional relaxation processes present at extremely low frequencies. Bannenberg \textit{et al.} \cite{bannenberg2016magnetic} also identified a low-frequency contribution to the dissipation in Fe$_{1-x}$Co$_x$Si that could not be described by the Cole-Cole model both at the conical-to-skyrmion and the skyrmion-to-conical transitions. 



\section{Conclusion}
In this study, we investigated the dynamics of the field-driven phase transitions between the magnetic states in GaV$_4$S$_8$ via ac susceptibility measurements. The magnetic response related to the continuous deformations of the cycloidal structure and the skyrmion lattice occurring inside the phases, such as the field-induced anharmonicity and transverse distortions, are accompanied with an instantaneous response of the spin system on the frequency scale well above the kHz range. The emergence of extremely slow dynamics was demonstrated at the phase boundaries between the cycloidal, skyrmion lattice and field-polarized states.

Similar frequency-dependent susceptibility and peaks in the imaginary (dissipative) part of the susceptibility characterize the phase boundaries between modulated magnetic states in all cubic helimagnets \cite{bauer2016generic}. However, in contrast to the lack of low-frequency dissipative processes near the conical-FP boundary in helimagnets, in GaV$_4$S$_8$, the Cyc-FP transition is characterized by extremely slow dynamical processes, possibly associated to the discontinuous transition into the FP state through the unwinding of the 2$\pi$ domain walls.

Magnetization dynamics at the Cyc-SkL and SkL-FP phase transitions in GaV$_4$S$_8$ is well described by the Cole-Cole relaxation model. However, discrepancies found at the Cyc-FP transition indicate the presence of more complex dynamics that cannot be described by a distribution of Debye-relaxation processes. 

Each field-driven phase transition in GaV$_4$S$_8$ shows similar behavior as a function of the temperature: at the high-temperature end of the phase boundaries the characteristic relaxation times are much shorter than 1\,ms, whereas at the lower-temperature part of the phase boundaries, the dynamics is drastically slowed down, characterized by relaxation times well above the minute scale. Similar temperature dependence of the relaxation times has been reported for Cu$_2$OSeO$_3$ \cite{levatic2014dissipation,qian2016phase} and may be a common feature for all bulk skyrmion host materials. The broad distribution of relaxation times and their strong dependence on the temperature, implying large activation energies, are likely the results of the collective relaxation of large magnetic structures of various volumes. 

The extremely slow relaxation times at the low-temperature end of the phase boundaries may be exploited for the creation of non-equilibrium skyrmionic clusters in the FP state with a long lifetime, whereas a relatively fast switching can be realized between the modulated states at higher temperatures.

\section{Acknowledgement}
The authors thank I. \v{Z}ivkovi{\'c} and H. R\o nnow for enlighting discussions. This work was supported by the Deutsche Forschungsgemeinschaft through the Transregional Collaborative Research Center TRR 80 and by the Hungarian Research Funds OTKA K 108918, OTKA PD 111756 and Bolyai 00565/14/11.

%

\end{document}